\documentclass[journal,article,submit,pdftex,oneauthor]{Definitions/mdpi}

\usepackage{xcolor}
\makeatletter

\let\linenumbers\relax

\makeatother

\newcommand{\rev}[1]{#1}
\newcommand{\revTwo}[1]{#1}
\newcommand{\revOLS}[1]{#1}
\firstpage{1} 
\pubvolume{1}
\issuenum{1}
\articlenumber{0}
\pubyear{2026}
\copyrightyear{2026}
\datereceived{ } 
\daterevised{ } 
\dateaccepted{ } 
\datepublished{ } 

\Title{Quantifying Perception-Based Student Success with Generative AI: An Exploratory Monte Carlo Simulation $^\dagger$ 
}

\Author{Seyma Yaman Kayadibi 
 \orcidA{}}

\AuthorNames{Seyma Yaman Kayadibi}

\address[1]{%
Institute for Sustainable Industries and Liveable Cities, Victoria University, Melbourne, VIC 3011, 
 Australia; seyma.yamankayadibi@live.vu.edu.au}

\corres{Correspondence: seyma.yamankayadibi@live.vu.edu.au}

\firstnote{\revTwo{This article is an extended and substantially revised version of a conference paper presented at the Melbourne Institute of Technology ICETE Conference, Sydney, NSW, Australia, 9--10 February 2026. The earlier conference version is available at https://doi.org/10.25397/ppny-f488.}}

\abstract{Generative artificial intelligence (GenAI) tools such as ChatGPT 
 have attracted growing attention in higher education, particularly in relation to how students perceive their usefulness, usability, and educational value. \rev{However, existing studies are often descriptive and rarely translate perception data into exploratory quantitative indicators that can support structured evaluation under uncertainty.} \rev{To address this gap, this study develops an exploratory Monte Carlo simulation framework for quantifying perception-based student success in the context of GenAI use.} \rev{The term Perception-Based Student Success Score is used here as an exploratory proxy indicator derived from students' positive evaluations of usability, efficiency, learnability, and perceived integration; it does not represent direct academic achievement, grades, retention, or objectively measured learning outcomes.} \rev{A PRISMA-informed structured literature search in Scopus identified nineteen empirical studies published between 2023 and 2025, of which six reported item-level means and standard deviations suitable for probabilistic modelling.} \revTwo{One coherent 10-item, 5-point Likert-scale usability-oriented instrument was selected as a canonical proof-of-concept dataset and used to parameterise an inverse-variance-weighted Monte Carlo simulation generating 10{,}000 synthetic observations.} The results show that the weighting structure substantially influences the simulated outcome. In particular, \textit{System Efficiency and Learning Burden} received the largest inverse-variance weight and therefore had the strongest influence on the composite score. \revTwo{This dominance should be interpreted cautiously because low variance in Likert-scale data may reflect response homogeneity or ceiling effects rather than substantive importance alone.} \revTwo{The study offers a transparent, reproducible, and privacy-preserving proof-of-concept framework linking structured literature search, item-level summary statistics, and probabilistic modelling.}}


\begin{document}
\nolinenumbers

\section{Introduction}

Generative artificial intelligence (GenAI), particularly large language models such as ChatGPT, has rapidly entered higher education and begun to reshape how students approach learning, writing, problem solving, information seeking, and academic support. Since its public release, ChatGPT has attracted substantial attention across universities, with widespread awareness and use reported across academic fields, gender groups, and levels of study. In this sense, GenAI has emerged not only as a practical educational resource but also as a continuing source of debate concerning pedagogy, assessment, academic integrity, and the future role of human judgment in learning environments (St\"ohr et al., 2024).

Recent empirical evidence suggests that student use of ChatGPT is already substantial across a wide range of educational settings. In Australia, more than one-third of undergraduate students were reported to use AI chatbots for assistance with assessments, often without perceiving such use as a violation of academic integrity (Gruenhagen et al., 2024). This finding is important because it shows that the spread of GenAI in higher education is not merely a matter of technological adoption; it also changes how students interpret legitimate support, authorship, and academic responsibility. Students in Hong Kong likewise described generative AI as useful for brainstorming, drafting, and idea development, while also raising concerns about reliability, transparency, and the need for clearer institutional guidance (Chan \& Hu, 2023). Together, these findings suggest that student engagement with GenAI is no longer peripheral, but is increasingly embedded in everyday academic practice.

At the same time, the motives for GenAI use and the consequences associated with such use vary considerably across contexts. One study found that academic workload and time pressure were positively associated with ChatGPT use among university students, suggesting that students often turn to such tools under demanding academic conditions (Abbas et al., 2024). The same study also indicated that while GenAI may help students cope with academic pressure, its use may coexist with less desirable outcomes such as procrastination and memory-related difficulties (Abbas et al., 2024). In language-learning settings, ChatGPT-assisted learning has been shown to be closely connected to motivational and self-regulatory factors, indicating that the value of GenAI may extend beyond convenience toward more personalised forms of academic support (Rahimi et al., 2025). In second-language writing contexts, students also evaluated ChatGPT in relation to teacher guidance, local infrastructure, and the practical demands of academic tasks (Meniado et al., 2024). These findings indicate that student use of GenAI is shaped not only by the availability of the technology itself, but also by workload pressures, learner motivation, academic strategy, and educational environment.

Student attitudes toward GenAI are likewise multidimensional rather than uniformly positive or negative. Students have been shown to evaluate ChatGPT through a combination of intention to use, verification of information, and responsible use, indicating that responses to GenAI frequently involve both perceived benefit and caution (Acosta Enriquez et al., 2024). In another context, students recognised the usefulness of ChatGPT while also expressing concern that excessive reliance on the tool could weaken deep thinking and increase academic integrity risks (Alghazo et al., 2025). Broader work on responsible GenAI education similarly emphasises that educational adoption should remain attentive to foresight, contextualisation, and the continuing importance of human pedagogical judgment (Azc\'arate, 2024). Taken together, these studies suggest that student attitudes toward GenAI are best understood as a balance among convenience, efficiency, trust, accountability, and educational risk rather than as simple approval or rejection.

Important differences also appear across disciplinary, regional, and contextual settings. Disciplinary background, in particular, appears to influence both familiarity with and acceptance of AI tools. For example, computer science students were found to be generally more supportive of AI integration than social science students, suggesting that prior exposure to digital systems and disciplinary norms may shape perceptions of educational usefulness (Dolenc \& Brumen, 2024). More broadly, students' use and perceptions of AI chatbots have been shown to vary across fields of study, academic levels, and demographic groups, while formal instruction on how such tools should be used has often remained limited (St\"ohr et al., 2024). In programming education, generally positive views of ChatGPT have also been reported across multiple African contexts, although students simultaneously raised concerns about cultural bias, inclusion, and broader educational implications (Oyelere \& Aruleba, 2025). These findings underscore that the meaning of GenAI in education is not fixed. Rather, it is mediated by disciplinary culture, national context, educational infrastructure, and the perceived fit between the technology and specific academic tasks.

The literature also shows that students evaluate GenAI not only in terms of general usefulness, but also in relation to learning support, performance implications, creativity, and broader educational benefit. In creative and design-related settings, students' responses to GenAI have been shaped by how these tools support ideation, usability, and discipline-specific expectations (Balt\`a-Salvador et al., 2025; Chellappa \& Luximon, 2024). In some cases, no statistically significant performance differences were found between AI-assisted and non-assisted groups, although prior familiarity with ChatGPT was associated with stronger outcomes (Balt\`a-Salvador et al., 2025). In programming contexts, ChatGPT-supported students showed more frequent debugging behaviour and somewhat improved performance, although these gains were not always large or statistically decisive (Sun et al., 2024). In other contexts, interactions with a generative AI chatbot were associated with more elaborated dialogue, higher perceived usefulness, stronger behavioural intention to use, and improved creative problem-solving performance (Song et al., 2025). Students using AI for writing tasks have also frequently reported both greater positivity toward the technology and continuing dissatisfaction with limitations related to tone, emotional resonance, and critical depth (Wang et al., 2024). Collectively, these studies suggest that the educational value of GenAI is task-dependent and cannot be inferred from a single domain of use alone.

Further variation is visible in relation to usability, critical engagement, and structural inequality. Some students report substantial use of ChatGPT for assignments and projects while simultaneously demonstrating limited critical engagement with AI-generated content, which points to ongoing challenges in digital literacy and evaluative judgment (Valova et al., 2024). In health sciences education, ChatGPT was rated somewhat more favourably than traditional digital tools in usability terms, yet qualitative findings still highlighted concerns about misinformation and unclear academic integrity boundaries (Veras et al., 2024). Design students also expressed mixed reactions, with some appreciating the novelty and assistance of the technology, and others raising concerns about creativity, motivation, and ethics (Chellappa \& Luximon, 2024). From a digital divide perspective, attitudes toward ChatGPT have additionally been shown to vary according to structural and demographic conditions, suggesting that access, preparedness, and perceived educational benefit are not distributed evenly across student populations (Zhang et al., 2025). This means that student perceptions of GenAI are shaped not only by the functionality of the tools themselves, but also by broader inequalities in educational access, institutional support, and contextual readiness.

\rev{Despite the rapid growth of this literature, an important methodological gap remains. Most existing studies report descriptive survey findings, thematic analyses, usability results, or context-specific reflections on student attitudes, but relatively few attempt to translate these perceptions into structured quantitative indicators or probabilistic models. As a result, the literature has become increasingly rich in descriptive insight while remaining limited in its ability to organise perception-based evidence into formal composite measures that can be examined transparently and reproducibly. This limitation becomes especially important when researchers wish to move beyond isolated survey findings and instead ask how multiple perception dimensions might be integrated into a broader exploratory evaluation framework. Methodological critiques point toward the need for stronger comparative, predictive, and quantitatively integrative designs capable of moving beyond descriptive reporting of student perceptions alone (Chellappa \& Luximon, 2024). Related work has also highlighted the need for stronger quantitative analysis of what students actually gain from AI-supported learning tasks (Wang et al., 2024). From a digital divide perspective, there is likewise a need for models that can compare student evaluations under uneven conditions of access and support (Zhang et al., 2025). In this respect, the field still lacks a formal framework for integrating diverse perception dimensions into an exploratory composite estimate of perceived success in using GenAI.}

\rev{The present study addresses this methodological gap by combining a PRISMA-informed structured literature search with Monte Carlo simulation. The literature search was used to identify empirical studies on student perceptions of GenAI in higher education and to determine which studies reported item-level summary statistics suitable for probabilistic modelling.} \revTwo{Through this process, 19 empirical studies were identified as relevant to the topic, but only six reported item-level means and standard deviations in a form suitable for simulation-based modelling.} \rev{These six studies therefore define the current quantitative boundary of the available evidence base for the proposed framework. Within this subset, one thematically coherent and statistically complete Likert-scale instrument was selected as a canonical proof-of-concept dataset for parameterising an inverse-variance-weighted Monte Carlo simulation. The remaining eligible datasets are treated as a natural extension point for future comparative modelling.} \revTwo{Accordingly, the structured literature search functions as a methodological filtering and contextualisation stage rather than as a full meta-analysis or exhaustive evidence synthesis.}

\rev{In higher education research, student success is widely understood as a multidimensional construct rather than a single outcome measure, encompassing dimensions such as academic achievement, satisfaction, persistence, skill development, and post-education outcomes (York et al., 2015). In educational technology and information systems research, success is also treated as a multidimensional construct that includes not only objective outcomes, but also users' perceived usefulness, ease of use, satisfaction, system quality, and perceived benefits (Davis, 1989; DeLone \& McLean, 2003). This broader understanding is relevant to GenAI in higher education because long-term academic outcome evidence remains limited, whereas student perception data are already widely reported.} \revTwo{In the present study, the term Perception-Based Student Success Score therefore describes an exploratory proxy indicator of perceived success in GenAI-supported learning activity.} \rev{It captures how positively students evaluate the tool in relation to usability, efficiency, learnability, and integration, while remaining distinct from direct measures of academic achievement such as grades, retention, or longitudinal learning gains.}

\rev{Building on this structured search process, the study develops a Monte Carlo-based framework in which perception-oriented analytical domains are translated into a probabilistic structure used to quantify a composite Perception-Based Student Success Score. In this study, ``success'' is defined operationally as perceived success in using GenAI tools in educational settings. This does not imply direct measurement of grades, retention, long-term academic achievement, or objectively observed learning outcomes. Instead, it refers to the way students positively evaluate their own experience of using GenAI as part of their learning process, particularly in relation to usability, efficiency, learnability, and perceived integration.} \revTwo{Such a framing is useful in the current state of the literature, where perception data are abundant but directly comparable quantitative outcome models remain scarce. By converting item-level summary statistics into simulated composite scores, the study shows how perception-based evidence can be analysed in a more structured, transparent, and uncertainty-aware manner.}

\rev{The use of the term Perception-Based Student Success Score is therefore intentionally limited. It should not be interpreted as a generalisable measure of academic success, nor as evidence that positive perceptions necessarily produce improved academic performance. Rather, it is used as an exploratory proxy indicator grounded in students' positive evaluations of GenAI-related usability, efficiency, learnability, and perceived integration. This distinction is important because perception-based measures may indicate whether a technology is experienced as supportive, usable, and beneficial, but they do not replace direct educational outcome measures.} \revTwo{The present framework therefore provides a methodological bridge between descriptive perception research and future outcome-based validation.}

\rev{Accordingly, this study addresses the following research question: \textit{How can Monte Carlo simulation based on item-level summary statistics identified through a structured literature search be used to quantify perception-based student success with generative AI in higher education?} The purpose of the study is to develop an exploratory probabilistic model of perception-based student success using published survey statistics on generative AI. More specifically, the study aims to provide a transparent and privacy-preserving quantitative framework for examining how core GenAI-related perception domains, such as usability, system efficiency, and perceived complexity, shape students' evaluations of their own success, while also establishing a foundation for future work that links these perceptions more directly to educational outcomes.}

\rev{The contribution of this paper is therefore both substantive and methodological. Substantively, it contextualises recent empirical evidence on student perceptions of GenAI in higher education across multiple contexts, disciplines, and forms of use. Methodologically, it demonstrates how published item-level perception statistics can be incorporated into an exploratory probabilistic model that captures both composite estimates and uncertainty. In doing so, the study offers a transparent, reproducible, and privacy-preserving proof-of-concept framework for examining how student perceptions of GenAI may be modelled quantitatively,} \revTwo{while acknowledging the limits of the current evidence base and the need for future multi-source validation.}

\section{Materials and Methods 
}

\subsection{Overview and Rationale}

\rev{This study employed a hybrid exploratory design consisting of a PRISMA-informed structured literature search followed by a Monte Carlo simulation framework.} The design was selected to address a methodological gap in the current literature on student perceptions of generative artificial intelligence (GenAI) in higher education. Although prior studies have examined dimensions such as usability, trust, affect, responsible use, and educational benefit, most have remained descriptive and have not translated survey-based perception data into a probabilistic modelling framework. \revTwo{In response, the present study combined structured evidence identification with simulation-based estimation to construct a Perception-Based Student Success Score.}

\rev{The term Perception-Based Student Success Score is used in a deliberately limited sense. It refers to an exploratory proxy indicator derived from students' positive evaluations of GenAI-related usability, efficiency, learnability, and perceived integration. It does not represent direct academic achievement, grades, retention, or objectively measured learning outcomes. Rather, it provides a structured way to examine how favourable perception-based indicators may be translated into a reproducible composite measure under uncertainty.}

The simulation framework was designed to model uncertainty explicitly rather than rely on a single-point weighted average. Using published item-level means and standard deviations, the model generated \rev{10{,}000 synthetic learner-level observations} and aggregated the resulting \rev{analytical domain scores} through inverse-variance weighting. This approach yields a full distribution of the composite score, propagates uncertainty from items to \rev{analytical domains} and from \rev{domains} to the overall score, and supports sensitivity-oriented interpretation when \rev{weighting assumptions} vary. Because the framework relies solely on published summary statistics, it is privacy-preserving, transparent, and reproducible. In this respect, the design is consistent with broader educational modelling approaches that use Monte Carlo procedures to examine outcomes under uncertainty (Torres et al., 2021).

\revTwo{The study should therefore be interpreted as a proof-of-concept simulation framework rather than as a full meta-analysis or exhaustive evidence synthesis.} \rev{The structured literature search was used to identify and contextualise relevant empirical studies and to determine which studies reported item-level summary statistics suitable for simulation-based modelling. The Monte Carlo component was then implemented using one canonical dataset that provided complete and coherent item-level information.} \revTwo{The methodological novelty of this approach is its adaptation of inverse-variance weighting and Monte Carlo simulation to published item-level perception statistics identified through a structured literature search, thereby producing a transparent, reproducible, and privacy-preserving framework for perception-based evaluation.}

\subsection{Structured Literature Search}

\rev{The search process was conducted as a PRISMA-informed structured literature search for the identification, screening, and reporting of studies relevant to student perceptions of GenAI in higher education (Page et al., 2021).} \rev{Scopus was selected as the sole database because the purpose of the search was not to produce the broadest possible thematic coverage or a full meta-analytic synthesis, but to identify empirical studies reporting item-level summary statistics suitable for simulation-based modelling. The use of a single database is acknowledged as a limitation and is treated as a future extension point, particularly through inclusion of additional databases such as Web of Science.} The initial search used the following query:

\begin{quote}
TITLE-ABS-KEY (``ChatGPT'' OR ``Generative AI'' OR ``Generative Artificial Intelligence'') AND TITLE-ABS-KEY (``students'' OR ``student'') AND (``perception'' OR ``performance'' OR ``trust'' OR ``learning'' OR ``success'').
\end{quote}

\rev{This search yielded 2{,}191 English-language, open-access records.} To focus more specifically on empirical higher education research, a second and more restrictive abstract-based query was applied:

\begin{quote}
ABS (``ChatGPT'' OR ``Generative AI'' OR ``Generative Artificial Intelligence'') AND ABS (``students'' OR ``university students'' OR ``higher education'') AND ABS (``perception'' OR ``attitude'' OR ``academic performance'' OR ``learning outcome'') AND ABS (``survey'' OR ``experiment'' OR ``empirical'' OR ``questionnaire'').
\end{quote}

This step reduced the pool to 274 articles. Manual relevance screening in Scopus yielded 49 core studies. Of these, 19 full-text empirical articles met the final inclusion criteria of examining student perceptions of GenAI in higher education. Among these 19 studies, six reported complete item-level means and standard deviations on 5-point Likert scales in a form suitable for probabilistic modelling (Chan \& Hu, 2023; Chellappa \& Luximon, 2024; Dolenc \& Brumen, 2024; Gruenhagen et al., 2024; Oyelere \& Aruleba, 2025; Veras et al., 2024). These six studies define the current quantitative boundary of the evidence base for the proposed framework.

Within this subset, the study by Veras et al.\ (2024) was selected as the \rev{canonical proof-of-concept dataset} for simulation because it reported a \rev{coherent 10-item, 5-point usability-oriented instrument} with complete item-level means and standard deviations. The remaining five item-complete datasets were treated as a natural extension point for future comparative modelling. Accordingly, the \rev{structured literature search served primarily as a methodological filtering and contextualisation stage for identifying statistically viable simulation inputs rather than as a broad narrative synthesis or integrated quantitative synthesis of all GenAI perception research.}

\revTwo{This design choice was intentional. The purpose of the simulation was to demonstrate the internal logic, transparency, and reproducibility of the modelling pipeline before extending it to a multi-source simulation framework. Therefore, the results should be understood as dataset-specific, exploratory, and methodologically demonstrative rather than as generalisable estimates derived from all nineteen studies.}

As shown in Figure~\ref{fig:searchflow}, the \rev{PRISMA-informed structured search process} progressively narrowed the initial Scopus results to a small subset of studies suitable for probabilistic modelling.

\begin{figure}[H]
\centering
\fbox{\parbox{0.72\textwidth}{\centering
Scopus search records identified\\
\textbf{(n = 2,191)}
}}
\vspace{0.4cm}
$\downarrow$
\vspace{0.4cm}
\fbox{\parbox{0.72\textwidth}{\centering
Records retained after restrictive abstract-based filtering\\
\textbf{(n = 274)}
}}
\vspace{0.4cm}
$\downarrow$
\vspace{0.4cm}
\fbox{\parbox{0.72\textwidth}{\centering
Core studies after manual relevance screening in Scopus\\
\textbf{(n = 49)}
}}
\vspace{0.4cm}
$\downarrow$
\vspace{0.4cm}
\fbox{\parbox{0.72\textwidth}{\centering
Full-text empirical studies meeting inclusion criteria\\
\textbf{(n = 19)}
}}
\vspace{0.4cm}
$\downarrow$
\vspace{0.4cm}
\fbox{\parbox{0.72\textwidth}{\centering
Studies with complete item-level means and standard deviations for modelling\\
\textbf{(n = 6)}
}}
\vspace{0.4cm}
$\downarrow$
\vspace{0.4cm}
\fbox{\parbox{0.72\textwidth}{\centering
Canonical proof-of-concept dataset selected for Monte Carlo simulation\\
\textbf{(n = 1)}
}}
\caption{\rev{PRISMA-informed 
 structured literature search flow diagram showing study identification, screening, eligibility, and final selection for probabilistic modelling.}}
\label{fig:searchflow}
\end{figure}
\subsection{Representative Study Selection and Simulation Preparation}

To operationalise the simulation framework, one representative study was selected from the final set of six studies that satisfied the Monte Carlo simulation criteria. \rev{Although all six studies reported compatible descriptive statistics based on 5-point Likert scales with item-level means and standard deviations, the present analysis intentionally uses one canonical proof-of-concept dataset rather than attempting a multi-source synthesis.} \revTwo{The study by Veras et al.\ (2024) was chosen as the canonical empirical input because it reported a coherent 10-item usability-oriented instrument with complete item-level statistics, making it especially suitable for demonstrating the modelling framework in detail.} In the original study, the instrument was used to examine students' experiences, perceptions, and usability judgments regarding ChatGPT in undergraduate health sciences education (Veras et al., 2024). \rev{Accordingly, the simulation should be interpreted as dataset-specific and methodologically demonstrative rather than as a generalisable synthesis of all studies identified through the structured literature search.}

For the purposes of the present analysis, the 10 questionnaire items were reorganised into three \rev{exploratory analytical themes}. \revTwo{These themes are conceptual and analytical groupings used to organise the item-level simulation inputs in a transparent and interpretable way; they are not presented as psychometrically validated latent factors.} Theme 1, \textit{Ease of Use and Learnability}, captured intuitive usability and users' confidence in operating the system. Theme 2, \textit{System Efficiency and Learning Burden}, captured the cognitive and technical effort associated with system use. Theme 3, \textit{Perceived Complexity and Integration}, captured perceived consistency, operational difficulty, and the extent to which the system's functions were experienced as integrated. This thematic organisation was informed by the wording and functional orientation of the selected usability-oriented items, particularly those related to ease of use, learnability, complexity, confidence, integration, and technical support requirements (Bangor et al., 2008; Brooke, 1996; Veras et al., 2024).

\rev{The three-theme structure was retained because it provided a conceptually interpretable way to distinguish between different aspects of students' usability-oriented perceptions. However, because the original SUS-based logic is commonly treated as broadly unidimensional, the present thematic organisation should be understood as an exploratory analytical structure rather than as a validated factor model. No exploratory or confirmatory factor analysis was conducted because the study relied on published item-level summary statistics rather than respondent-level raw data.} Rather than collapsing all usability-related perceptions into a single undifferentiated construct, this structure allowed the simulation framework to distinguish between intuitive use, efficiency-related burden, and perceived system coherence in a way that remained consistent with usability-oriented interpretation (Bangor et al., 2008; Brooke, 1996).

More specifically, Theme 1 comprised Q1 (``use regularly''), Q3 (``easy to use''), Q7 (``most would quickly learn''), and Q9 (``high confidence''). Theme 2 comprised three negatively worded items: Q2 (``too complicated''), Q4 (``need technical assistance''), and Q10 (``had to learn much''). Theme 3 comprised Q5 (``functions integrated well''), Q6 (``too many inconsistencies''), and Q8 (``difficult to operate'').

Negatively worded items in Themes 2 and 3 were reverse-coded so that higher values consistently represented more favourable perceptions. This \rev{exploratory thematic structure} provided the analytical basis for the inverse-variance aggregation of item scores into theme-level composites and, subsequently, for the Monte Carlo simulation of the composite \revTwo{Perception-Based Student Success Score}. For notational clarity, \(i\) indexes items within a theme, \(k\) indexes themes, \(t\) indexes synthetic \rev{learner-level observations}, and \(L\) denotes the Likert maximum, with \(L = 5\). The total number of simulated \rev{observations} is denoted by \(N = 10{,}000\). The allocation of items to themes is summarised in Table~\ref{tab:theme_mapping}.

\begin{table}[H]
\caption{\rev{Exploratory 
 allocation of questionnaire items to the three analytical themes used in the simulation framework.}}
\label{tab:theme_mapping}
\centering
\begin{tabularx}{\textwidth}{>{\raggedright\arraybackslash}p{3.4cm} >{\raggedright\arraybackslash}p{1.2cm} X}
\toprule
\textbf{Theme} & \textbf{Item} & \textbf{Item content} \\
\midrule
\multirow{4}{3.4cm}{Ease of Use and Learnability} 
& Q1 & I believe that I would like to use this system regularly. \\
& Q3 & I found it easy to use the system. \\
& Q7 & I would guess that most would quickly learn to use the system. \\
& Q9 & I felt highly confident operating with this system. \\
\midrule
\multirow{3}{3.4cm}{System Efficiency and Learning Burden} 
& Q2 & The system is too complicated, in my opinion. (reverse-coded) \\
& Q4 & I would require the assistance of someone technical. (reverse-coded) \\
& Q10 & Before I could get going, I had to learn much. (reverse-coded) \\
\midrule
\multirow{3}{3.4cm}{Perceived Complexity and Integration} 
& Q5 & The functions within the system were integrated well. \\
& Q6 & There were too many inconsistencies in this system. (reverse-coded) \\
& Q8 & It was quite difficult to operate such a system. (reverse-coded) \\
\bottomrule
\end{tabularx}
\end{table}

\subsection{Reverse-Coding Procedure}

Items in Themes 2 and 3 that were negatively worded were reverse-coded so that higher values uniformly indicated more favourable perceptions. This procedure is consistent with standard usability scale practice, including the logic commonly applied in the System Usability Scale literature (Bangor et al., 2008; Brooke, 1996). For a 5-point Likert scale, reverse-coding was defined as 

\begin{equation}
x' = L + 1 - x,
\end{equation}

where \(x\) is the original response and \(L\) is the Likert maximum. Reverse-coding aligned item directionality, prevented distortion in aggregated means and variances, and ensured that all items could be aggregated on a common directional scale within each theme.

\subsection{Weighting Procedure}

For each item \(i\), an inverse-variance weight \(w_i\) was computed from the reported standard deviation \(s_i\) as

\begin{equation}
w_i = \frac{1}{s_i^2}.
\end{equation}

\rev{In this study, inverse-variance weighting is used as an exploratory weighting mechanism rather than as a formal meta-analytic estimator. In conventional meta-analysis, inverse-variance weighting is commonly used to combine estimates across studies. Here, by contrast, it is applied within a single canonical dataset to item-level standard deviations in order to construct variance-sensitive thematic composites.} Let \(\mu_i\) denote the reported mean of item \(i\). Theme-level means were then calculated as normalised weighted averages:

\begin{equation}
\bar{x}_{\mathrm{theme}} = \frac{\sum_i w_i \mu_i}{\sum_i w_i}.
\end{equation}

This procedure preserves the relative contribution of items with lower reported dispersion and yields a \rev{variance-sensitive thematic composite}. \revTwo{However, in Likert-scale data, lower variance does not necessarily imply greater substantive importance or measurement precision; it may also reflect response homogeneity, ceiling effects, or restricted scale use. Therefore, the dominance of any theme under inverse-variance weighting is interpreted as a feature of the modelling framework rather than as definitive evidence of substantive causal importance.}

\subsection{Theme-Level Variance Estimation}

To estimate theme-level variance while retaining inverse-variance weights in unnormalised form, a Bessel-type correction was applied \revTwo{to reduce} variance underestimation in small composite scales. \rev{This correction was used as part of the exploratory item-aggregation framework and should not be interpreted as reconstructing respondent-level variance from raw data.} Let \(M\) denote the number of items in a given theme, \(w_i\) the inverse-variance weight of item \(i\), \(\mu_i\) its mean, and \(\bar{x}_{\mathrm{theme}}\) the weighted theme mean. The corrected weighted variance was calculated as

\begin{equation}
s_{\mathrm{theme}}^2 =
\frac{M}{M-1}
\cdot
\frac{\sum_i w_i (\mu_i - \bar{x}_{\mathrm{theme}})^2}{\sum_i w_i}.
\end{equation}

\rev{When all weights are equal, this expression corresponds to a Bessel-corrected dispersion measure over item-level means.} In the present framework, it yields \rev{a corrected weighted dispersion estimate} that can be carried forward into the simulation stage. \revTwo{Accordingly, the theme-level variance is treated as a weighted dispersion measure derived from item means within each theme, rather than as respondent-level variance reconstructed from individual participant responses. The Monte Carlo procedure should therefore be interpreted as a summary-statistics-based synthetic modelling approach.}

\subsection{Monte Carlo Simulation Approach}

\rev{Perception-based student success was modelled using a Monte Carlo procedure that transformed item-level Likert summary statistics into synthetic \textit{Perception-Based Student Success Scores}.} For each theme, the item-level means and corrected variances were first aggregated into theme-level parameters. Then, for each synthetic \rev{learner-level observation} \(t\) and each theme \(k \in \{1,2,3\}\), a latent theme score was drawn from a normal distribution:

\begin{equation}
T_{t,k} \sim \mathcal{N}(\mu_k, s_k^2),
\end{equation}

where \(\mu_k\) and \(s_k^2\) denote the inverse-variance weighted theme mean and corrected theme variance, respectively. \rev{The final composite scores were constrained to the 1--5 interval in order to preserve the bounds of the original Likert metric.} \revTwo{This normal approximation is a simplifying modelling assumption. Because Likert-scale responses are discrete and bounded, the simulated values should not be interpreted as reconstructed individual responses, but as summary-statistics-based synthetic approximations used to examine how the proposed scoring framework behaves under uncertainty.}

Each synthetic \rev{learner-level observation's} composite Perception-Based Student Success Score was then computed as an inverse-variance weighted average of the three theme scores with an additive Gaussian noise term:

\begin{equation}
S_t^* = \frac{\sum_k w_k T_{t,k}}{\sum_k w_k} + \varepsilon_t,
\end{equation}

where \(w_k = 1/s_k^2\) is the inverse of the corrected theme-level variance estimate and \(\varepsilon_t \sim \mathcal{N}(0, 0.05^2)\) represents a small stochastic perturbation introduced to avoid treating the composite score as a deterministic transformation of the thematic components. The resulting values \(S_t^*\) were then clipped to the 1--5 range, yielding the final simulated Perception-Based Student Success Score \(S_t\) for each of the 10{,}000 synthetic \rev{learner-level observations}.

This framework does more than reproduce a single weighted average. It yields the full distribution of the composite score, propagates uncertainty from items to themes and from themes to the final outcome, and allows future scenario analysis through alternative weighting assumptions. Because the simulation uses only published summary statistics, it is especially suitable in contexts where individual-level microdata are unavailable but item-level means and standard deviations are reported. \revTwo{The implementation remains specific to the selected canonical dataset and should be interpreted as a proof-of-concept simulation rather than as a reconstruction of the original respondent-level data or a generalisable estimate across all higher education contexts.} More broadly, this use of Monte Carlo simulation is consistent with educational modelling under uncertainty, while the present implementation remains specific to the structure of the selected GenAI perception data (Torres et al., 2021; Veras et al., 2024).

\section{Results}

\subsection{Theme-Level Composite Estimates}

Following the inverse-variance weighting and Bessel-corrected variance procedures described in the previous section, weighted means and standard deviations were obtained for each of the three \rev{exploratory thematic composites}. These theme-level estimates served as the empirical inputs for the Monte Carlo simulation and summarised the central tendency and dispersion of the canonical usability-based dataset on the original 1--5 Likert scale. \rev{Because the thematic structure is exploratory and not psychometrically validated as a factor model, the results should be interpreted as descriptive inputs to the simulation framework rather than as estimates of latent constructs.}

For Theme 1, \textit{Ease of Use and Learnability}, four items (Q1, Q3, Q7, and Q9) were included. No reverse-coding was required. Using the reported item means of 3.71, 4.21, 4.33, and 4.00, together with item standard deviations of 0.75, 0.66, 0.56, and 0.83, the inverse-variance weighted mean was estimated as \(\bar{x}_1 \approx 4.12\). The corresponding Bessel-corrected weighted standard deviation was approximately \(SD_1 \approx 0.27\).

For Theme 2, \textit{System Efficiency and Learning Burden}, three negatively worded items (Q2, Q4, and Q10) were reverse-coded so that higher values indicated more favourable perceptions. Using the reversed item means of 4.08, 4.25, and 4.08 and standard deviations of 0.58, 0.79, and 0.78, the inverse-variance weighted mean was estimated as \(\bar{x}_2 \approx 4.12\). The Bessel-corrected weighted standard deviation for this theme was substantially smaller, at approximately \(SD_2 \approx 0.09\), \rev{which gave this theme the largest inverse-variance weight within the proposed modelling framework.}

For Theme 3, \textit{Perceived Complexity and Integration}, three items (Q5, Q6, and Q8) were included, with Q6 and Q8 reverse-coded. Using the reversed item means of 3.88, 3.46, and 3.62 and standard deviations of 0.61, 0.88, and 0.82, the inverse-variance weighted mean was estimated as \(\bar{x}_3 \approx 3.71\). The corresponding Bessel-corrected weighted standard deviation was approximately \(SD_3 \approx 0.22\).

Taken together, these estimates show that Themes 1 and 2 had similarly high weighted means, whereas Theme 3 was lower. However, Theme 2 exhibited the smallest variance and therefore received the largest inverse-variance weight under the adopted weighting scheme. \revTwo{In Likert-scale data, smaller variance may reflect response homogeneity, ceiling effects, or restricted use of the scale rather than greater substantive importance alone. Accordingly, the prominence of Theme 2 is treated as a feature of the weighting-based simulation framework, not as evidence of an independent causal relationship.}

\subsection{Simulated Perception-Based Student Success Score Distribution}

Using the theme-level parameters reported above, a Monte Carlo simulation was implemented in Python 3.10.11 using the Thonny IDE 
 to generate a synthetic sample of \(N = 10{,}000\) \rev{learner-level observations}. A reproducible Python implementation of the main Monte Carlo simulation is provided in Appendix~A. For each simulated \rev{observation}, theme-level scores were drawn from the corresponding weighted normal distributions and combined into a composite Perception-Based Student Success Score according to the inverse-variance weighting procedure defined in Equation~(6). The resulting values were clipped to the 1--5 interval to preserve the scale of the original Likert instrument. \revTwo{Because the simulation is based on published summary statistics rather than raw respondent-level data, the simulated observations should be interpreted as synthetic modelling outputs rather than as reconstructed student responses.}

\begin{figure}[H]
\centering
\includegraphics[width=0.75\textwidth]{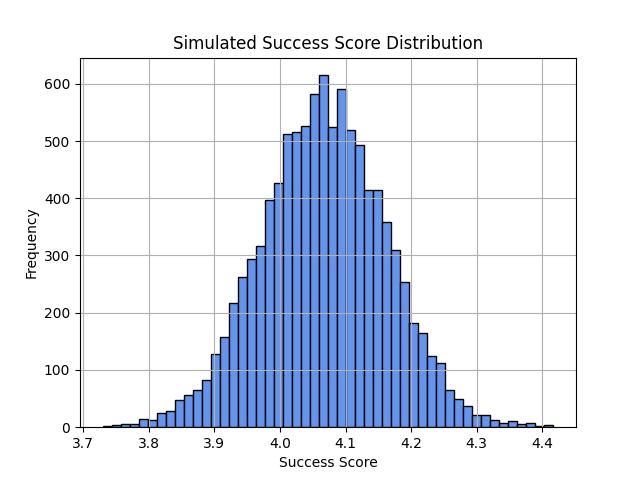}
\caption{\rev{Histogram of simulated \textit{Perception-Based Student Success Scores} (\(N = 10{,}000\)). The distribution is concentrated around a high mean, indicating generally favourable perception-based evaluations of GenAI use within the selected canonical dataset.}}
\label{fig:histogram}
\end{figure}

The simulated Perception-Based Student Success Score distribution was tightly concentrated around a high central value, indicating generally favourable perception-based evaluations of GenAI use within the selected canonical dataset. The mean simulated Perception-Based Student Success Score was 4.0666, with a standard deviation of 0.0956. The interquartile range extended from 4.0032 to 4.1308, showing that the majority of simulated values were closely clustered around the mean. A fixed random seed was used in the Python implementation to ensure full reproducibility of the simulated sample. The resulting distribution served as the basis for the histogram presented in Figure~\ref{fig:histogram}. \revTwo{These values are best interpreted as exploratory outputs of the proposed proof-of-concept simulation framework rather than as generalisable estimates of student success across higher education.}

\subsection{\rev{Internal Diagnostic Check of Theme Contributions}}

\revTwo{An internal diagnostic check was conducted to examine whether the simulated Perception-Based Student Success Score reflected the weighting structure used in its construction. Because \(S_t\) was constructed directly from the same thematic inputs \(T_{t,1}\), \(T_{t,2}\), and \(T_{t,3}\), this analysis is interpreted only as a diagnostic check, not as an independent explanatory, predictive, or causal model.}

\rev{The fitted diagnostic equation was}

\begin{equation}
S_t = -0.0248 + 0.0856T_{t,1} + 0.7823T_{t,2} + 0.1381T_{t,3}.
\end{equation}

\revTwo{The coefficient for Theme 2 was the largest because Theme 2 had the smallest corrected variance and therefore received the largest inverse-variance weight in the score construction. Theme 1 and Theme 3 also showed positive but smaller coefficients. These coefficients confirm the internal mechanics of the simulated score rather than showing that one thematic construct independently predicts student success.}

\revTwo{The full OLS output is provided in Appendix~C for transparency. Inferential statistics such as \(p\)-values, \(F\)-statistics, residual diagnostics, and \(R^2\) are not treated as substantive evidence because the outcome was generated from the same inputs used as predictors.}
\subsection{\rev{Summary of Diagnostic and Simulation Findings}}

\revTwo{Overall, the simulated Perception-Based Student Success Score reflected the weighting structure defined in the modelling framework. All three themes contributed positively to the composite score by construction, but Theme 2, \textit{System Efficiency and Learning Burden}, had the largest influence because it had the smallest corrected variance and therefore received the largest inverse-variance weight. Themes 1 and 3 also contributed positively, although their relative influence was smaller under the adopted weighting scheme.}

\revTwo{Taken together, these findings indicate that, within the present proof-of-concept simulation framework, the composite score was most strongly shaped by the theme with the lowest corrected variance. This confirms how the inverse-variance weighting structure affects the simulated Perception-Based Student Success Score, rather than showing that system efficiency and cognitive burden independently predict student success. At the same time, the high mean of the simulated score indicates that the overall perception-based evaluative profile remained favourable within the selected canonical dataset. Because the diagnostic regression was fitted to a simulated outcome constructed from the same thematic inputs, the results are interpreted as confirmation of the internal logic of the weighting and simulation framework rather than as evidence of an independent causal or predictive relationship.}

\subsection{Sensitivity Analysis of Weighting Schemes}

To examine whether the prominence of the three thematic dimensions depended critically on the specific weighting structure, a sensitivity analysis was conducted across three alternative weighting schemes. The analysis was implemented in Python using the same simulated theme scores generated from the empirical means and standard deviations of Theme 1 (\textit{Ease of Use and Learnability}), Theme 2 (\textit{System Efficiency and Learning Burden}), and Theme 3 (\textit{Perceived Complexity and Integration}). By holding the simulated theme scores constant and varying only the weighting structure used to construct the composite Perception-Based Student Success Score, the analysis isolated the effect of the weighting scheme itself from random variation in the simulated inputs.

Three scenarios were examined. The first used the baseline inverse-variance weighting scheme adopted in the main analysis, in which each theme received a weight proportional to the inverse of its empirical variance. The second used equal weights across all three themes, thereby treating the composite Perception-Based Student Success Score as a simple unweighted average. The third used a reduced-gap weighting scheme, constructed as the midpoint between the baseline inverse-variance vector and the equal-weights vector. This third specification retained the same ordering of theme weights as the baseline model but reduced the disparity between the largest and smallest weights.

For each scenario, the composite Perception-Based Student Success Score was recomputed for \(N = 10{,}000\) simulated \rev{learner-level observations} by combining the same theme scores with the scenario-specific weights and a small Gaussian noise term (\(\sigma = 0.05\)). The resulting scores were then clipped to the 1--5 interval. \revTwo{For transparency, the same internal diagnostic regression form, \(S_t \sim T_{t,1} + T_{t,2} + T_{t,3}\), was fitted within each scenario; these regressions are interpreted only as checks of how the alternative weighting schemes are reflected in the simulated score, not as independent predictive models.} A reproducible Python implementation of the weighting-scheme sensitivity analysis is provided in Appendix~B.

Under the baseline inverse-variance specification, the diagnostic coefficients reproduced the main internal check, with \(\beta_0 \approx -0.0248\), \(\beta_1 \approx 0.0856\), \(\beta_2 \approx 0.7823\), and \(\beta_3 \approx 0.1381\), and with \(R^2 \approx 0.7240\). Under the equal-weights specification, the coefficients became nearly symmetric, with \(\beta_0 \approx 0.0107\), \(\beta_1 \approx 0.3312\), \(\beta_2 \approx 0.3315\), and \(\beta_3 \approx 0.3350\), while \(R^2 \approx 0.8535\). This pattern is expected because the Perception-Based Student Success Score in that scenario was constructed as an almost uniform average of the three themes. Under the reduced-gap specification, the coefficients remained close to the baseline values, with \(\beta_0 \approx 0.0214\), \(\beta_1 \approx 0.0879\), \(\beta_2 \approx 0.7665\), and \(\beta_3 \approx 0.1404\), and with \(R^2 \approx 0.7175\).

\revTwo{Taken together, these results show that the simulated prominence of Theme 2 depends on the inverse-variance weighting structure and remains visible under the reduced-gap specification. However, because this weighting scheme gives greater influence to themes with lower corrected variance, and because low variance in Likert-scale data may reflect response homogeneity, ceiling effects, or restricted scale use, the finding should not be interpreted as evidence that Theme 2 has independent substantive or causal priority. The sensitivity analysis is therefore treated as a robustness check on the behaviour of the proposed scoring framework under alternative weighting assumptions. The baseline inverse-variance specification remains the primary modelling scenario, while the alternative schemes illustrate how the composite score changes when the weighting assumptions are relaxed.}

\section{Discussion}

\subsection{Interpretation of the Main Findings}

This study developed a privacy-preserving Monte Carlo framework for quantifying \textit{Perception-Based Student Success Scores} with generative artificial intelligence (GenAI) in higher education using published item-level summary statistics. \rev{The framework should be interpreted as an exploratory proof-of-concept model rather than as a direct measure of academic achievement or a generalisable estimate of student success across higher education.} The results indicate that the simulated Perception-Based Student Success Score was concentrated at the upper end of the 1--5 scale, suggesting generally favourable perception-based evaluations of GenAI use within the selected canonical dataset. At the same time, the three thematic dimensions did not contribute equally to the composite outcome. Under the baseline inverse-variance weighting scheme, \textit{System Efficiency and Learning Burden} received the largest weight, whereas \textit{Ease of Use and Learnability} and \textit{Perceived Complexity and Integration} also contributed positively but to a lesser extent.

\revTwo{This pattern is meaningful within the internal logic of the proposed modelling framework, but it should not be overinterpreted as an independent substantive or causal finding.} The result suggests that, under the adopted inverse-variance weighting procedure, the simulated score is strongly influenced by the theme with the smallest corrected variance. \revTwo{Within the present proof-of-concept model, perceptions related to system efficiency and learning burden therefore had the greatest influence on the simulated composite score under the adopted weighting scheme.} This interpretation is broadly consistent with studies showing that students often value GenAI for efficiency, assistance, and practical academic support rather than for novelty alone (Abbas et al., 2024; Gruenhagen et al., 2024; Veras et al., 2024). \revTwo{However, because the analysis uses Likert-scale summary statistics, lower variance may also reflect response homogeneity, ceiling effects, or restricted scale use. The prominence of \textit{System Efficiency and Learning Burden} should therefore be interpreted as a feature of the weighting-based simulation rather than as definitive evidence that this theme is inherently more important than the others.}

The weighting-scheme sensitivity analysis further clarifies the behaviour of the model. Although the equal-weights specification mechanically forced all three themes to contribute almost symmetrically, the reduced-gap scenario preserved the general baseline pattern, with \textit{System Efficiency and Learning Burden} remaining the largest contributor under the modified weighting structure. \revTwo{This suggests that the prominence of Theme 2 is not solely dependent on the exact baseline weight values, although it remains conditional on the broader inverse-variance logic used in the model.} In this respect, the analysis extends simulation-based approaches used in educational decision modelling under uncertainty to the GenAI context (Torres et al., 2021), while remaining specific to the selected dataset and modelling assumptions.

These findings should nevertheless be interpreted with care. The present framework models perceived success rather than directly observed academic achievement. Accordingly, the Perception-Based Student Success Score should be understood as an exploratory perception-based proxy, not as a direct measure of grades, retention, or long-term learning outcomes. \rev{The simulated sample should likewise be interpreted as a set of synthetic learner-level observations generated from published summary statistics, not as reconstructed respondent-level data.} Similarly, the regression results are diagnostic rather than causal, because the simulated outcome was constructed from the same thematic inputs later entered into the regression model. The role of the regression is therefore to confirm the internal coherence of the simulation and weighting scheme rather than to establish an independent predictive relationship.

For interpretive benchmarking only, the mean simulated Perception-Based Student Success Score may also be linearly rescaled from the 1--5 range to a 0--100 SUS-equivalent index using

\begin{equation}
\mathrm{SUS}^{(\mathrm{eq})} = 100 \cdot \frac{S - 1}{4}.
\end{equation}

Using \(S \approx 4.07\), this yields \(\mathrm{SUS}^{(\mathrm{eq})} \approx 76.75\). This transformation is heuristic and should not be interpreted as a re-scoring of the original System Usability Scale items. Rather, it offers a familiar interpretive reference point for readers accustomed to usability-oriented benchmarking (Bangor et al., 2008; Brooke, 1996). \revTwo{Because the present score is a simulated perception-based composite rather than the original SUS score, the SUS-equivalent index is reported only as an approximate interpretive aid.}

\subsection{Contributions to Literature and Practice}

This study makes both conceptual and methodological contributions to the growing literature on generative artificial intelligence in higher education. Methodologically, it demonstrates how published item-level summary statistics can be transformed into an exploratory Perception-Based Student Success Score through inverse-variance weighting and Monte Carlo simulation. In doing so, it moves beyond descriptive summaries and provides a structured proof-of-concept framework capable of generating not only a point estimate but also a full simulated distribution of perception-based evaluation under uncertainty. This is especially relevant in educational contexts where raw participant-level data are unavailable or inaccessible for ethical, privacy, or institutional reasons (Borenstein et al., 2009; Torres et al., 2021).

\revTwo{The study also contributes conceptually by offering a cautious operationalisation of perceived success in GenAI-supported learning. Rather than treating success as directly observed academic achievement, the framework defines it as an exploratory perception-based proxy derived from favourable evaluations of usability, efficiency, learnability, and perceived integration. This distinction is important because the present model does not measure grades, retention, or longitudinal learning gains; instead, it shows how perception-based evidence can be organised into a transparent simulation framework for later validation against direct educational outcomes.}

The study further organises the selected usability-oriented instrument into three \rev{exploratory analytical themes}: \textit{Ease of Use and Learnability}, \textit{System Efficiency and Learning Burden}, and \textit{Perceived Complexity and Integration}. \rev{This structure is derived from a single canonical instrument and is not presented as a psychometrically validated factor model.} Nevertheless, it offers a coherent and empirically interpretable basis for demonstrating the modelling pipeline. In this sense, the framework provides a reusable analytical procedure that can be extended to additional instruments, themes, and contexts as more suitable item-level summary statistics become available.

From a practical standpoint, the findings suggest that educational institutions and system designers may benefit from attending to cognitive and operational burden when evaluating or implementing GenAI tools. \revTwo{Within the present modelling framework, \textit{System Efficiency and Learning Burden} received the largest inverse-variance weight and therefore had the strongest influence on the simulated composite score. This does not establish that the theme independently predicts student success; rather, it indicates that efficiency- and burden-related perceptions are important candidates for future empirical validation.} Tools may be valued less for novelty alone and more for their capacity to simplify academic tasks, save time, and support efficient workflows. \rev{Accordingly, the proposed Perception-Based Student Success Score should be understood as an exploratory evaluation tool rather than as a definitive benchmark for deployment readiness or institutional decision-making.}

\subsection{Implications for Practice and Policy}

The practical implications of the present findings are closely aligned with the weighting structure observed in the model. Most importantly, the prominence of \textit{System Efficiency and Learning Burden} suggests that, within the selected canonical dataset and adopted weighting scheme, students' favourable evaluations of GenAI use were closely associated with perceptions of reduced cognitive and operational burden. \revTwo{This finding should be interpreted cautiously because the theme's prominence is partly a consequence of the inverse-variance weighting procedure and may also reflect response homogeneity or ceiling effects in Likert-scale data.} Nevertheless, it supports the practical relevance of workflow-centred design. Institutions should therefore consider whether GenAI tools reduce friction, simplify routine academic tasks, and fit naturally into students' existing practices rather than focusing only on technological novelty or advanced functionality. Related studies in programming and AI-supported learning likewise suggest that students often value GenAI when it supports practical academic processes and is perceived as facilitating task completion (Oyelere \& Aruleba, 2025; Sun et al., 2024).

Integration quality also remains important. Although \textit{Perceived Complexity and Integration} contributed less strongly than Theme 2 under the baseline weighting structure, its positive contribution indicates that smooth embedding within institutional digital environments may still matter. GenAI tools that are difficult to navigate, poorly aligned with learning management systems, or inconsistent in operation may weaken students' perceptions of educational benefit even when their core functionality is strong. Related empirical work also indicates that students' perceptions vary across fields of study and educational contexts, suggesting that institutional implementation should remain sensitive to disciplinary differences rather than assume that one model of GenAI adoption will fit all settings (Dolenc \& Brumen, 2024; St\"ohr et al., 2024).

Policy implications should therefore be framed cautiously. \revTwo{The present study does not provide direct policy evidence based on observed academic outcomes; rather, it offers an exploratory modelling framework for identifying perception-based dimensions that may require closer institutional attention.} Higher education institutions should develop clearer, evidence-informed guidance on acceptable use, information verification, ethical use, and academic integrity rather than relying on broad or ambiguous digital adoption strategies. Students have reported both perceived benefits and concerns in relation to GenAI, including support for learning, idea generation, and efficiency alongside concerns about ethics, accuracy, and policy ambiguity (Acosta Enriquez et al., 2024; Chan \& Hu, 2023; Gruenhagen et al., 2024; Veras et al., 2024). This suggests that effective policy should combine technical adoption with pedagogical guidance and institutional clarity.

Issues of access and equity should also remain central. Digital divide perspectives suggest that institutional planning for GenAI should take account of uneven access, unequal support, and context-specific constraints rather than assuming uniform readiness across student populations (Oyelere \& Aruleba, 2025; Zhang et al., 2025). The portability of the present simulation framework is especially relevant in such settings because it allows researchers to explore perception-based outcomes using published summary statistics without requiring access to sensitive individual-level data. \revTwo{However, institutional use of the framework should remain exploratory until the score is tested across multiple datasets and validated against direct educational outcomes.}

\subsection{Limitations and Future Research}

Several limitations should be acknowledged. First, although 19 empirical studies met the broader eligibility criteria, only six reported item-level means and standard deviations in a form suitable for probabilistic modelling. This constrains the quantitative breadth of the present framework and reflects a wider reporting limitation in the current literature. \revTwo{In addition, the search was limited to Scopus. The search process should therefore be understood as a PRISMA-informed structured literature search rather than as exhaustive evidence synthesis or full meta-analysis. Future research should extend the search strategy to additional databases, including Web of Science, to increase coverage and improve the comprehensiveness of the evidence base.}

Second, the thematic operationalisation used in the simulation was derived from a single canonical proof-of-concept dataset, namely Veras et al.\ (2024), which necessarily introduces the possibility of instrument-specific bias. The present analysis should therefore be interpreted as a canonical proof-of-concept application rather than a comprehensive synthesis of all available perception studies. \revTwo{Although six studies reported item-level summary statistics, the simulation was intentionally restricted to one coherent usability-oriented instrument to demonstrate the internal logic and reproducibility of the modelling pipeline before extension to a multi-source framework. This limits generalisability and means that the results should be interpreted as dataset-specific rather than as estimates derived from all studies identified through the structured literature search.}

\revTwo{Third, the three-theme structure used in the simulation should be interpreted as an exploratory analytical organisation rather than as a psychometrically validated factor structure. The selected instrument is usability-oriented, and the original SUS-based logic is commonly treated as broadly unidimensional. Because the present study relied on published item-level summary statistics rather than respondent-level raw data, exploratory or confirmatory factor analysis could not be conducted. Future studies using raw participant-level data should test whether the proposed thematic organisation is empirically supported.}

\revTwo{Fourth, the inverse-variance weighting procedure should be interpreted cautiously. In conventional meta-analysis, inverse-variance weighting is typically used to combine estimates across studies. In the present study, it was adapted as an exploratory within-dataset weighting mechanism based on item-level standard deviations. For Likert-scale data, lower variance does not necessarily indicate greater measurement precision or substantive importance; it may also reflect response homogeneity, ceiling effects, or restricted use of the response scale. Therefore, the dominance of \textit{System Efficiency and Learning Burden} under the baseline weighting scheme should be interpreted as a feature of the modelling framework rather than as definitive evidence of substantive priority.}

\revTwo{Fifth, the Monte Carlo simulation used a bounded normal approximation for Likert-scale summary statistics. Although simulated values were constrained to the 1--5 interval, Likert responses are discrete and ordinal, and normal approximations may be imperfect, particularly near the upper end of the scale. Future work should compare this approach with alternative distributional assumptions, including ordinal simulation strategies, truncated-normal specifications, beta-binomial approximations, or empirical resampling where raw data are available.}

Sixth, the current model does not yet incorporate potentially important moderators such as discipline, prior exposure to GenAI, task type, level of study, or institutional context, despite clear evidence that perceptions and usage differ across such dimensions (Dolenc \& Brumen, 2024; St\"ohr et al., 2024). Seventh, while the study introduced a weighting-scheme sensitivity analysis, it has not yet expanded to broader forms of robustness testing such as systematically varying the noise variance, exploring alternative distributional assumptions, or conducting external validation against observed academic outcomes. These steps remain necessary if the framework is to be evaluated more fully as a predictive or benchmarking tool.

\revTwo{Finally, the regression analysis reported in this study should be interpreted only as an internal diagnostic check. Because the simulated Perception-Based Student Success Score was constructed from the same thematic inputs used in the regression, the regression does not provide independent explanatory, predictive, or causal evidence. Its purpose is limited to confirming that the simulated score reflects the weighting structure used to construct it.}

Future research should therefore extend the model in several directions. First, multiple eligible instruments should be incorporated to increase thematic robustness and reduce dependence on a single canonical dataset. Second, future simulation work should broaden the sensitivity analysis to include alternative assumptions about noise, distributions, and weighting structures. Third, and most importantly, the Perception-Based Student Success Score should be validated longitudinally against observed academic outcomes such as performance, persistence, or learning gains. Such extensions would help determine whether the present framework can function not only as a perception-based proxy but also as a useful comparative indicator for evidence-based decision-making in higher education.

Despite these limitations, the present study provides a replicable and privacy-preserving first step toward scalable quantitative evaluation of GenAI in higher education. \revTwo{Its contribution is primarily methodological: it shows how published item-level perception statistics can be transformed into an interpretable simulation framework while making the assumptions and limits of that transformation explicit.} By transforming item-level summary statistics into an uncertainty-aware simulation framework, it opens a path toward more cumulative, comparative, and cautious analysis of student perceptions in the rapidly evolving landscape of AI-supported learning.

\section{Conclusions}

\rev{This study showed how item-level summary statistics identified through a PRISMA-informed structured literature search can be translated into an exploratory and uncertainty-aware Monte Carlo simulation framework for quantifying perception-based student success with generative artificial intelligence (GenAI) in higher education.} Of the 49 core studies screened in detail, 19 empirical studies met the final eligibility criteria, but only six reported item-level means and standard deviations in a form suitable for probabilistic modelling. Within this quantitative boundary, the instrument reported by Veras et al.\ (2024) provided a coherent canonical dataset for simulation and enabled the operationalisation of a three-theme exploratory analytic framework based on \textit{Ease of Use and Learnability}, \textit{System Efficiency and Learning Burden}, and \textit{Perceived Complexity and Integration}.

Using inverse-variance weighting and Monte Carlo simulation, the framework generated a synthetic sample of \rev{10{,}000 learner-level observations} and produced a composite Perception-Based Student Success Score on the original 1--5 scale. The mean simulated Perception-Based Student Success Score was 4.0666, with a standard deviation of 0.0956, indicating that the simulated distribution was concentrated toward the favourable end of the scale \rev{within the selected canonical dataset}. At the theme level, \textit{Ease of Use and Learnability} and \textit{System Efficiency and Learning Burden} both produced weighted means of approximately 4.12, whereas \textit{Perceived Complexity and Integration} was lower at approximately 3.71. However, \textit{System Efficiency and Learning Burden} showed the smallest variance and therefore received the largest inverse-variance weight under the adopted weighting scheme.

\revTwo{The internal diagnostic analysis confirmed that the simulated score reflected the weighting structure used to construct it. Under the baseline specification, \textit{System Efficiency and Learning Burden} showed the largest coefficient because it had the smallest corrected variance and therefore received the largest inverse-variance weight. This result confirms the internal mechanics of the simulation and weighting procedure rather than showing that this theme independently predicts student success.} A weighting-scheme sensitivity analysis further showed how the simulated composite score changed under baseline inverse-variance, equal-weight, and reduced-gap weighting scenarios. \revTwo{These sensitivity checks support the transparency of the modelling framework while also indicating the need for external validation using additional datasets and observed educational outcomes.}

Methodologically, the study contributes a portable and privacy-preserving simulation pipeline that operates entirely on published summary statistics. This is especially valuable in higher education settings where participant-level data may be inaccessible because of privacy, governance, or ethical constraints. In this respect, the study extends simulation-based educational modelling into the GenAI context and demonstrates how perception-based evidence can be organised into a structured quantitative framework rather than remaining limited to descriptive survey summaries (Torres et al., 2021; Veras et al., 2024). \revTwo{The contribution should therefore be understood primarily as methodological and exploratory, not as a definitive measurement model of academic success.}

At the same time, the study should be interpreted within clear limits. The quantitative evidence base remains small, the search was limited to Scopus, the thematic structure was operationalised from a single canonical proof-of-concept dataset, and the framework has not yet been validated against observed academic outcomes such as grades, retention, or longitudinal learning gains. \revTwo{In addition, the simulation relies on a bounded normal approximation for Likert-scale summary statistics, and the inverse-variance weighting scheme may give greater influence to themes with lower variance because of response homogeneity or ceiling effects.} Although the weighting-scheme sensitivity analysis provided an initial robustness check, broader validation remains necessary through alternative distributional assumptions, expanded moderator structures, and testing across multiple datasets and institutional contexts.

Overall, this study demonstrates that Monte Carlo simulation based on published item-level perception statistics can be used to quantify a Perception-Based Student Success Score with GenAI in higher education in a way that is transparent, reproducible, and theoretically interpretable. Rather than replacing direct educational outcome measures, the proposed score should be understood as a structured exploratory proxy that may support future comparative evaluation after further validation. Future research should extend the framework across multiple eligible instruments, incorporate moderators such as discipline and task type, expand robustness analyses, and undertake longitudinal validation against observed academic outcomes in order to strengthen both generalisability and interpretive confidence.

\vspace{6pt}


\funding{This research received no external funding.}

\institutionalreview{Not applicable.}

\informedconsent{Not applicable.}

\dataavailability{\revTwo{The study used published summary statistics extracted from studies identified through the PRISMA-informed structured literature search. No new participant-level data were collected. The synthetic observations generated in the Monte Carlo simulation are reproducible from the parameters and Python code reported in the article. All data supporting the reported results are contained within the article and its referenced sources.}}

\acknowledgments{This research was supported by an Australian Government Research Training Program (RTP) Scholarship awarded to the author.}

\conflictsofinterest{The author declares no conflicts of interest.}
\abbreviations{Abbreviations}{
The following abbreviations are used in this manuscript:
\\

\noindent
\begin{tabular}{@{}ll}
GenAI & Generative artificial intelligence\\
SUS & System Usability Scale
\end{tabular}
}

\appendixtitles{yes}
\appendixstart
\appendix

\section{Python Implementation of the Monte Carlo Simulation}

\rev{The Python code used to generate the simulated theme scores, compute the composite Perception-Based Student Success Score, estimate descriptive statistics, and fit the internal diagnostic regression model is reproduced below.}

{\ttfamily\scriptsize
\noindent

import numpy as np\\
import pandas as pd\\
import statsmodels.api as sm\\
import matplotlib.pyplot as plt\\[4pt]

\# 1. Simulation settings\\
np.random.seed(42)\\
n = 10000  \# number of synthetic Monte Carlo observations\\[4pt]

\# 2. Simulate theme scores based on published summary statistics\\
T1 = np.random.normal(4.1169, 0.2709, n)\\
T2 = np.random.normal(4.1240, 0.0910, n)\\
T3 = np.random.normal(3.7100, 0.2160, n)\\[4pt]

\# 3. Inverse-variance weights\\
w1 = 1 / (0.2709 ** 2)\\
w2 = 1 / (0.0910 ** 2)\\
w3 = 1 / (0.2160 ** 2)\\
W\_total = w1 + w2 + w3\\[4pt]

\# 4. Perception-Based Student Success Score with small stochastic perturbation\\
random\_error = np.random.normal(0, 0.05, n)\\
Success = (w1 * T1 + w2 * T2 + w3 * T3) / W\_total \\
\hspace*{1em}+ random\_error\\
Success = np.clip(Success, 1, 5)\\[4pt]

\# 5. Create DataFrame\\
df = pd.DataFrame(\{\\
\hspace*{1em}"Theme1": T1,\\
\hspace*{1em}"Theme2": T2,\\
\hspace*{1em}"Theme3": T3,\\
\hspace*{1em}"Success\_Score": Success\\
\})\\[4pt]

\# 6. Descriptive statistics\\
print("Descriptive Statistics of Success Score:\textbackslash n")\\
print(df["Success\_Score"].describe())\\[4pt]

\# 7. Run multiple linear regression\\
X = sm.add\_constant(df[["Theme1", "Theme2", "Theme3"]])\\
y = df["Success\_Score"]\\
model = sm.OLS(y, X).fit()\\
print("\textbackslash nRegression Summary:\textbackslash n")\\
print(model.summary())\\[4pt]

\# 8. Optional: Visualize the distribution of Success Score\\
plt.hist(df["Success\_Score"], bins=50,\\
\hspace*{1em}color="cornflowerblue", edgecolor="black")\\
plt.title("Simulated Success Score Distribution")\\
plt.xlabel("Success Score")\\
plt.ylabel("Frequency")\\
plt.grid(True)\\
plt.show()
}

\section{Sensitivity Analysis of Weighting Schemes}

The Python code used to evaluate the baseline inverse-variance, equal-weights, and reduced-gap weighting scenarios is reproduced below.

{\ttfamily\scriptsize
\noindent
import numpy as np\\
import statsmodels.api as sm\\[4pt]

\# Sensitivity analysis over three weighting scenarios\\[4pt]

\# 1. Simulation settings\\
np.random.seed(42)\\
n = 10000  \# number of synthetic Monte Carlo observations\\[4pt]

\# 2. Simulate theme scores based on published summary statistics\\
T1 = np.random.normal(4.1169, 0.2709, n)\\
T2 = np.random.normal(4.1240, 0.0910, n)\\
T3 = np.random.normal(3.7100, 0.2160, n)\\[4pt]

\# 3. Baseline inverse-variance weights\\
w1 = 1 / (0.2709 ** 2)\\
w2 = 1 / (0.0910 ** 2)\\
w3 = 1 / (0.2160 ** 2)\\[4pt]

baseline\_weights = np.array([w1, w2, w3])\\
equal\_weights = np.array([1.0, 1.0, 1.0])\\
shrunk\_weights = 0.5 * baseline\_weights \\
\hspace*{1em}+ 0.5 * equal\_weights\\[4pt]

scenarios = [\\
\hspace*{1em}\{"name": "baseline\_inverse\_variance",\\
\hspace*{2em}"weights": baseline\_weights, "noise\_sd": 0.05\},\\
\hspace*{1em}\{"name": "equal\_weights",\\
\hspace*{2em}"weights": equal\_weights, "noise\_sd": 0.05\},\\
\hspace*{1em}\{"name": "reduced\_gap\_weights",\\
\hspace*{2em}"weights": shrunk\_weights, "noise\_sd": 0.05\},\\
]\\[4pt]

\# 4. Run sensitivity analysis over weighting schemes\\
for s in scenarios:\\
\hspace*{1em}w = s["weights"]\\
\hspace*{1em}W\_total = w.sum()\\[4pt]

\hspace*{1em}\# Success Score under given weights + small Gaussian perturbation\\
\hspace*{1em}noise = np.random.normal(0, s["noise\_sd"], n)\\
\hspace*{1em}Success = (w[0] * T1 + w[1] * T2 + w[2] * T3) / W\_total \\
\hspace*{2em}+ noise\\
\hspace*{1em}Success = np.clip(Success, 1, 5)\\[4pt]

\hspace*{1em}\# Multiple linear regression: Success \textasciitilde{} T1 + T2 + T3\\
\hspace*{1em}X = sm.add\_constant(np.column\_stack([T1, T2, T3]))\\
\hspace*{1em}y = Success\\
\hspace*{1em}model = sm.OLS(y, X).fit()\\[4pt]

\hspace*{1em}print(f"\textbackslash nScenario: \{s['name']\}")\\
\hspace*{1em}print("Coefficients (const, Theme1, Theme2, Theme3):")\\
\hspace*{1em}print(model.params)\\
\hspace*{1em}print("R-squared:", model.rsquared)
}

\section{\rev{Internal Diagnostic Regression Output}}
\label{app:diagnostic_regression}

\rev{The ordinary least squares (OLS) output below is provided only as an internal diagnostic check of the simulated Perception-Based Student Success Score. It is not interpreted as an independent explanatory, predictive, or causal regression model because the dependent variable was constructed directly from the same thematic inputs used as predictors. The purpose of including this output is limited to transparency and reproducibility.}

\begin{figure}[H]
\centering
\includegraphics[width=0.95\textwidth]{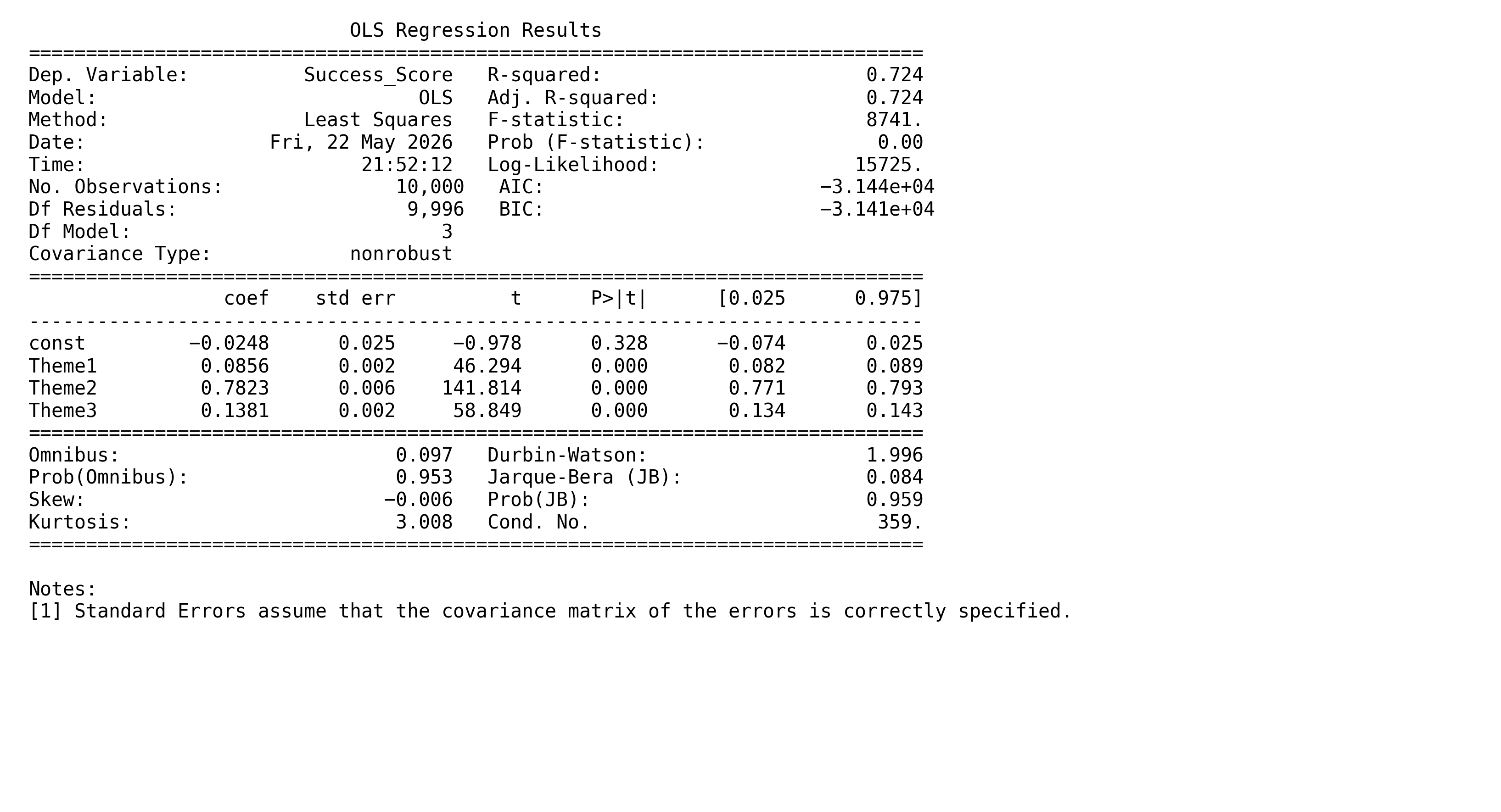}
\caption{\revOLS{The ordinary least squares (OLS)} 
 output for the internal diagnostic check of the simulated Perception-Based Student Success Score. The regression is included for transparency only and is not interpreted as evidence of independent predictive or causal relationships.}
\label{fig:regression_summary_appendix}
\end{figure}

\reftitle{References}

\PublishersNote{}

\end{document}